\documentclass[pre,aps,twocolumn,showpacs]{revtex4}
\usepackage{epsfig}
\include{graphics}
\begin{document}

\input{epsf}

\title{Hydrodynamics of fluid-solid coexistence in dense shear granular flow}

\author{Evgeniy Khain}

\affiliation{Department of Physics and Michigan Center for
Theoretical Physics, The University of Michigan, Ann Arbor,
Michigan 48109}

\begin{abstract}
We consider dense rapid shear flow of inelastically colliding hard
disks. Navier-Stokes granular hydrodynamics is applied accounting
for the recent finding that shear viscosity diverges at a lower
density than the rest of constitutive relations. New interpolation
formulas for constitutive relations between dilute and dense cases
are proposed and justified in molecular dynamics (MD) simulations.
A linear stability analysis of the uniform shear flow is performed
and the full phase diagram is presented. It is shown that when the
inelasticity of particle collision becomes large enough, the
uniform sheared flow gives way to a two-phase flow, where a dense
"solid-like" striped cluster is surrounded by two fluid layers.
The results of the analysis are verified in event-driven MD
simulations, and a good agreement is observed.
\end{abstract}
 \maketitle

\section{Introduction}

Granular shear flows have attracted much attention in recent years
\cite{attention}. However, theoretical description of {\it dense}
granular flows still remains a challenge \cite{Aranson}. The
present work focuses on a {\it rapid dense} flow of inelastic hard
disks, which undergo a shear motion. Here, the medium is
fluidized, particle collisions are binary and instantaneous. This
brings about a possibility of a hydrodynamic description of
granular media \cite{hydrreview}.

There are still significant difficulties in hydrodynamic
description of {\it dense} shear flows, even within the model of
inelastic hard spheres. Although for low and moderate densities, a
hydrodynamic description with the proper constitutive relations
can be derived from kinetic theories \cite{Haff}, constitutive
relations for dense flows are presently unknown. Several attempts
to extend the theory to high densities have been done recently
\cite{Grossman,Luding01,Meerson,Luding}. It was proposed to apply
free volume arguments \cite{freevolume,Grossman} in the vicinity
of close packing density, $n_{max}=2/(\sqrt{3} d^2)$, and use
interpolation functions for constitutive relations between low and
high density cases \cite{Grossman,Luding01,Meerson,Luding}. At
high densities, there is another problem related to the behavior
of shear viscosity \cite{Bocquet,Luding,Khain}. While inelastic
heat losses, thermal conductivity, and pressure diverge at the
close packing density $n_{max}$, it was shown recently that shear
viscosity diverges at a lower density \cite{Luding}. This may
result in coexistence of fluid and solid phases \cite{Khain}.

We propose here novel interpolation formulas for constitutive
relations between low-density and high-density cases, which
account for viscosity divergence, and justify them in MD
simulations. We employ the resulting granular hydrodynamics for
\emph{quantitative} description of fluid-solid coexistence in
shear granular flow and verify the hydrodynamic predictions in MD
simulations.

\section{The model and governing hydrodynamic equations}

Consider a plane Couette geometry: an ensemble of inelastically
colliding hard disks with unit masses and diameter $d$ is driven,
at zero gravity, by two walls. The top wall, located at $y = H/2$,
moves in the $x$-direction with velocity $u_0$, the bottom wall
moves in the opposite direction with the same velocity. The only
parameter that characterizes the inelasticity of collisions is the
coefficient of normal restitution, $r$. Boundary conditions in the
$x$-direction are periodic. In the $y$-direction, no-flux and
no-slip boundary conditions are implemented. Upon a collision with
a driving wall, the normal particle velocity switches sign, while
the new tangential velocity component is taken from
Maxwell-Boltzmann distribution, with a mean equal to the wall
velocity, and a variance corresponding to an instant temperature
of the layer next to the wall.

Now we present a hydrodynamic approach, which deals with the
number density of grains, $n({\mathbf r},t)$, the granular
temperature $T({\mathbf r},t)$, and the mean flow velocity
\cite{Haff}:
\begin{eqnarray} &&d n/d t + n\, {\mathbf \nabla} \cdot
{\mathbf v} = 0\,,\nonumber
\\
&&n\, (d{\mathbf v}/d t) = {\mathbf \nabla} \cdot {\mathbf P} \,,
\nonumber
\\
&&n\, (d T/d t) = - {\mathbf \nabla} \cdot {\mathbf Q} + {\mathbf
P}\,: {\mathbf \nabla} {\mathbf v} - \Gamma\,, \label{hydr1}
\end{eqnarray}
where ${\mathbf P}$ is the stress tensor, ${\mathbf Q}$ is the
heat flux, and $\Gamma$ is the energy losses due to the
inelasticity of particle collisions. The stress tensor ${\mathbf
P}$ is given by ${\mathbf P} =\left[ - p(n,T) + \mu(n,T)\,{\rm
tr}\, ({\mathbf D}\,) \right] {\mathbf I} + 2\eta(n,T)\,
\hat{{\mathbf D}}$, where ${\mathbf D} = (1/2)\left[{\mathbf
\nabla v} + ({\mathbf \nabla v})^{T}\right]$ is the rate of
deformation tensor, $\hat{{\mathbf D}}={\mathbf D}-\frac{1}{2}\,
{\rm tr}\, ({\mathbf D}\,)\, {\mathbf I}$ is the deviatoric part
of ${\mathbf D}$, and ${\mathbf I}$ is the identity tensor,
$\eta(n,T)$ and $\mu(n,T)$ are the shear (first) and bulk (second)
viscosities. The heat flux ${\mathbf Q}$ is given by ${\mathbf Q}
= -\kappa(n,T)\,{\mathbf \nabla T}$, where $\kappa(n,T)$ is the
coefficient of thermal conductivity. An additional term in the
expression for heat flux, proportional to the \emph{density}
gradient \cite{inelastic,Fourier}, can be neglected in the nearly
elastic limit $1-r^2 \ll 1$ we consider throughout this paper. For
higher values of inelasticity one also needs to take into account
inelastic corrections to transport coefficients \cite{inelastic}.
However, the main point of the present paper is addressing the
challenging problem of describing a very dense flow and (as the
first step) it is sufficient to assume the limit of nearly elastic
particle collisions.

\section{Testing constitutive relations in MD simulations}

For dilute and moderately dense granular flows, the Enskog-like
constitutive relations are derived from kinetic theory
\cite{Haff}. The shear viscosity, $\eta_E(n,T)$, the thermal
conductivity $\kappa_E(n,T)$, the inelastic heat losses
$\Gamma_E(n,T)$, and the equation of state $p_E$ are given by
\begin{eqnarray}
\eta_E &=& \frac{4\,\nu\,T^{1/2}\,G_E}{\pi^{3/2}\,d}
\left[1+\frac{\pi}{8}\left(1+\frac{1}{G_E}\right)^2\right]\,,\nonumber
\\
\kappa_E &=& \frac{8 \nu\,T^{1/2}\,G_E}{\pi^{3/2}\,d}
\left[1+\frac{9\pi}{16}\left(1+\frac{2}{3\,G_E}
\right)^2\right]\,,\nonumber
\\
\Gamma_E &=&\frac{8(1-r)n\,T^{3/2}\,G_E}{\pi^{1/2}\,d}\,,\nonumber
\\
p_E&=&n\,T(1+2\,G_E)\,, \label{Enskog}
\end{eqnarray}
where $G_E=\nu(1-7\nu/16)/(1-\nu)^2$ and $\nu=n\,(\pi\, d\,^2/4)$
is the solid fraction, and index $E$ stays for "Enskog". For small
and moderate densities and for small inelasticities, transport
coefficients are in a reasonable agreement with the results of
event-driven MD simulations \cite{Bizon}. For very large
densities, one can use free volume arguments
\cite{freevolume,Grossman} and show that all the constitutive
relations (except for shear viscosity) diverge at the density of
close packing. However, there is strong evidence that at high
densities the coefficient of shear viscosity behaves differently
than other constitutive relations \cite{Bocquet,Luding}, diverging
at a \emph{lower} density than other transport coefficients
\cite{Luding}. An immediate consequence of this finding is a
possible existence of a solid-like phase, which is at rest or
moves as a whole, as its density is higher than the density of
viscosity divergence \cite{Khain}. Khain and Meerson \cite{Khain}
considered a very dense three-dimensional system and assumed a
leading order expansion of constitutive relations (except for
shear viscosity) near the close packing density. However, to
employ hydrodynamic equations within a wide range of densities,
one needs a pragmatic approach involving interpolation of
constitutive relations between the dilute and dense cases
\cite{Grossman,Luding01,Meerson,Luding}. We will adapt this
approach, suggesting several new interpolation functions, and
testing them in MD simulations.

\begin{figure}[ht]
%\vspace{-0.3cm}
%\centerline{\includegraphics[width=7.0cm,clip=]{new_heatlosses1.eps}}
\includegraphics[width=6cm,clip=]{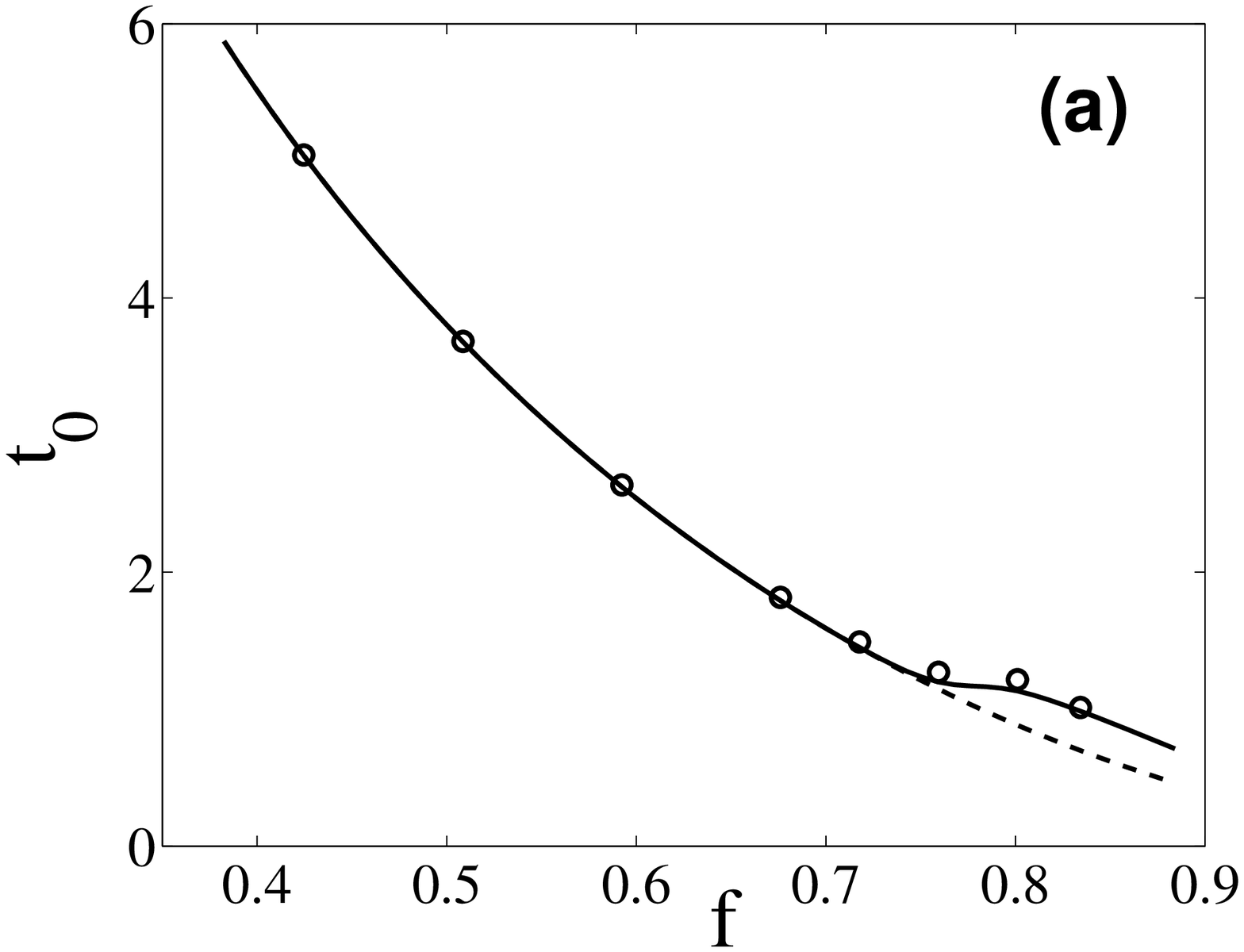}
\includegraphics[width=6cm,clip=]{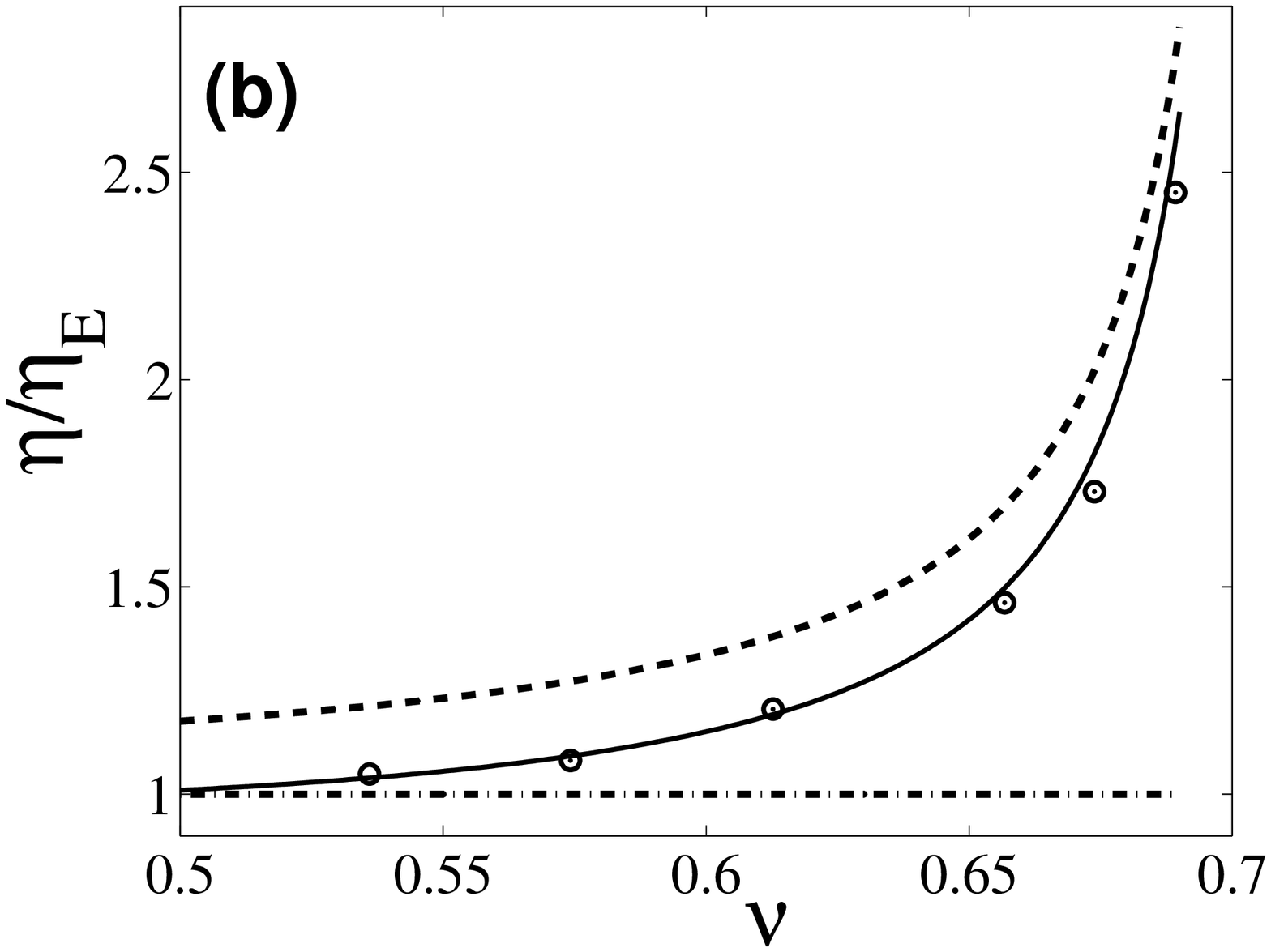}
\caption{Testing constitutive relations in MD simulations: heat
losses (a) and shear viscosity (b). (a): A time scale for a
temperature decay, $t_0$, as a function of density both from
theory (solid line) and MD simulations (circles). The Enskog
predictions (dashed line) do not agree with the results of MD
simulations at high densities. The restitution coefficient is
$r=0.999$ (the simulation with the highest density was performed
for $r=0.9999$ and then the value of $t_0$ was properly
recalculated), the total number of particles is $N = 6480$, and
the system aspect ratio is $\delta = 0.8$. (b): The coefficient of
shear viscosity normalized by the Enskog value as a function of
density. Enskog formula (dash-dotted line) disagrees completely
with MD simulations (circles) at high densities. Also shown are
the expression in Ref. \cite{Luding} (dashed line) and the new
interpolation formula (solid line). The total number of particles
is $N \approx 6400$, and the restitution coefficient is very close
to unity (ranging from $r=0.996$ to $r=0.99995$ in different MD
simulations), to ensure that the solution of uniform shear flow is
realized.} \label{constitutive}
\end{figure}

To begin with, a global equation of state of hard disk fluid was
recently proposed \cite{Luding01,Luding}: $p = n\,T(1+2\,G)$,
where $G = G_E + m \,( \nu_{max} / ( \nu_{max} - \nu  ) - G_E )$.
Here, $m$ is an interpolation function, given by $m = [ 1+\exp(
(\nu_c - \nu )/ m_0 ) ]^{-1}$ with $\nu_c = 0.70$, $m_0 = 0.0111$,
and $\nu_{max} = \pi/(2\sqrt{3})$, in a good agreement with MD
simulations \cite{Luding01}. According to the same lines, we
propose the modified form of inelastic heat losses, changing $G_E$
to $G$ in the corresponding expression in Eqs.~(\ref{Enskog}). To
test this expression in MD simulations, we consider a completely
different system: a homogeneous freely cooling ensemble of
inelastic hard disks in a box. In this case the third of
Eqs.~(\ref{hydr1}) becomes $n\,\partial T/\partial t = -\Gamma$,
which gives the well-known Haff's law \cite{Haff}: $T = (1 +
t/t_0)^{-2}$. Here temperature is measured in units of the initial
temperature $T_0$, and time in units of $H\,T_0^{-1/2}$, where $H$
is the system width. In our case, $t_0 =
(\pi^{1/2}/4)\,(d/H)\,[(1-r)G]^{-1}$. We measured the temperature
as a function of time in MD simulations within a wide range of
densities, while other parameters were kept constant. To ensure
the homogeneous cooling state being stable \cite{freecooling}, we
took a sufficiently large restitution coefficient. As expected, we
found the Haff's law to be valid and computed the value of $t_0$.
Figure~\ref{constitutive}a shows $t_0$ versus $f=<\nu>/\nu_{max}$,
both from the theory (solid line) and MD simulations (circles). A
good agreement is observed, in contrast to the Enskog predictions
(dashed line), which deviate from the results of MD simulations at
high densities.

The next step is specifying the coefficient of shear viscosity.
Consider a dense uniform shear flow. Here, the energy balance
equation reduces to $\eta (du/dy)^2 = \Gamma$. Measuring the
temperature of the system in MD simulations at different
densities, we calculate the inelastic heat losses, and compute the
shear viscosity $\eta$. In Ref. \cite{Luding}, the coefficient of
shear viscosity of a system of elastic hard disks was measured
using the Helfand-Einstein expressions. It was found that $\eta$
diverges like $a_{\eta}\,(\nu_{\eta}-\nu)^{-1}$, where the
viscosity divergence density is $\nu_{\eta} = 0.71$, and $a_{\eta}
= 0.037$ \cite{Luding}, in contrast to the Enskog predictions. Our
MD simulations confirm this result. Figure~\ref{constitutive}b
shows that the Enskog formula (dash-dotted line) dramatically
disagrees with MD simulations (circles) at high densities. An
interpolation between dilute and dense regime was also proposed in
Ref. \cite{Luding}: $\eta_L
=\eta_E\,[1+a_{\eta}/(\nu_{\eta}-\nu)]$, see
Fig.~\ref{constitutive}b, the dashed line. Since this
interpolation is not accurate for the intermediate densities, we
propose an improved interpolation function $\eta
=\eta_E\,[1+a_{\eta}(\nu/\nu_{\eta})^3/(\nu_{\eta}-\nu)-a_{\eta}/\nu_{\eta}]$.
The new formula (solid line) agrees well with MD simulations in a
wide range of densities.

In Ref. \cite{Luding}, the coefficient of thermal conductivity
$\kappa$ was also measured in MD simulations. It was found to be
larger than $\kappa_E$ for intermediate densities ($\nu \simeq
0.55$), and smaller than $\kappa_E$ for larger densities, $\nu
\simeq 0.75$ (see Fig. 7 in Ref. \cite{Luding}). Thermal
conductivity is known to diverge at the close packing density, as
$(\nu_{max} - \nu)^{-1}$. Incorporating these findings, we propose
the following interpolation formula $\kappa = \kappa_E\,(1 +
0.1\nu -10\nu^{10}+0.11(\nu_{max} - \nu)^{-1}-0.11/\nu_{max})$,
which is in a good agreement with the results of Ref.
\cite{Luding}. Now, we apply the resulting granular hydrodynamics
to the problem of dense shear flow.

\section{Steady dense shear flow: MD simulations versus
hydrodynamics}

Let us measure the coordinate $y$ in units of the system height
$H$, the horizontal velocity $u$ in units of the wall velocity
$u_0$, the temperature $T$ in units of $u^2_0$, the density $n$ in
units of $n_{max}$, the pressure $P$ in units of $n_{max}\,u^2_0$.
Then the steady shear flow ($\partial/\partial t = 0, v = 0,
\partial/\partial x = 0$) is described by the following system of
equations:
\begin{eqnarray}
\frac{d}{dy}\left(f_2T^{1/2}\frac{du}{dy} \right) = 0
\,,\,\,\,\,\,\,\,\,\,\,\,f_4 T=\mbox{const}\,,\nonumber
\\
\frac{d}{dy}\left(f_1 T^{1/2}\frac{dT}{dy} \right) + \frac{f_2
T^{1/2}}{4}\left(\frac{du}{dy}\right)^2 - R f_3 T^{3/2} = 0\,,
\label{steady}
\end{eqnarray}
where $R = (16/\pi)\,(1-r)\,(H/d)^2$ is the heat loss parameter,
and the functions $f_i$ are the density-dependent parts of the
constitutive relations: $f_1 = \kappa\,(\sqrt{\pi}d/2)T^{-1/2}$,
$f_2 = \eta\,(2d\sqrt{\pi})T^{-1/2}$, $f_3 = \nu\,G$, and $f_4 =
n(1+2G)$. No-flux and no-slip boundary conditions are given by
$dT/dy(y=-1/2) = dT/dy(y=1/2) = 0$, $u(y=-1/2) = -1$, $u(y=1/2) =
1$. The total number of particles is conserved, which yields a
normalization condition for the density: $\int_{-1/2}^{1/2}
n(y)\,dy = f$, where $f=<\nu>/\nu_{max}$ is the average area
fraction. Equations (\ref{steady}) allow for different solutions.
The simplest one is the uniform shear flow. In this case, density
and temperature are constant, and the velocity profile is linear.
For low and moderate densities, it is known that uniform shear
flow becomes unstable, when the inelasticity of particle
collisions exceeds a critical value
\cite{instability1,instability2,instability}. This problem has
been analyzed by linear stability analysis of the corresponding
equations of granular hydrodynamics
\cite{instability1,instability2,instability}. The instability may
result in shear-band formation, which was recently observed in MD
simulations \cite{Luding03} and analyzed theoretically for
moderate densities \cite{instability2}, see also
\cite{instability}. We found that for high densities, the same
instability exists, but its threshold changes qualitatively.
Indeed, as a result of shear viscosity divergence at
$\nu_{\eta}<\nu_{max}$, the uniform shear flow solution is
impossible at sufficiently large average density, $f
> \nu_{\eta}/\nu_{max}$. When this solution does not exist or is unstable,
the density profile is no more homogeneous. In this case, the
regions with the density larger than that of viscosity divergence
may appear. Although, those regions are at rest or move as a
whole, the granulate is fluidized there and granular temperature
is not zero \cite{Khain}.

\begin{figure}[ht]
\vspace{-0.3cm}
%\hspace{-0.1cm}
\includegraphics[width=6cm,clip=]{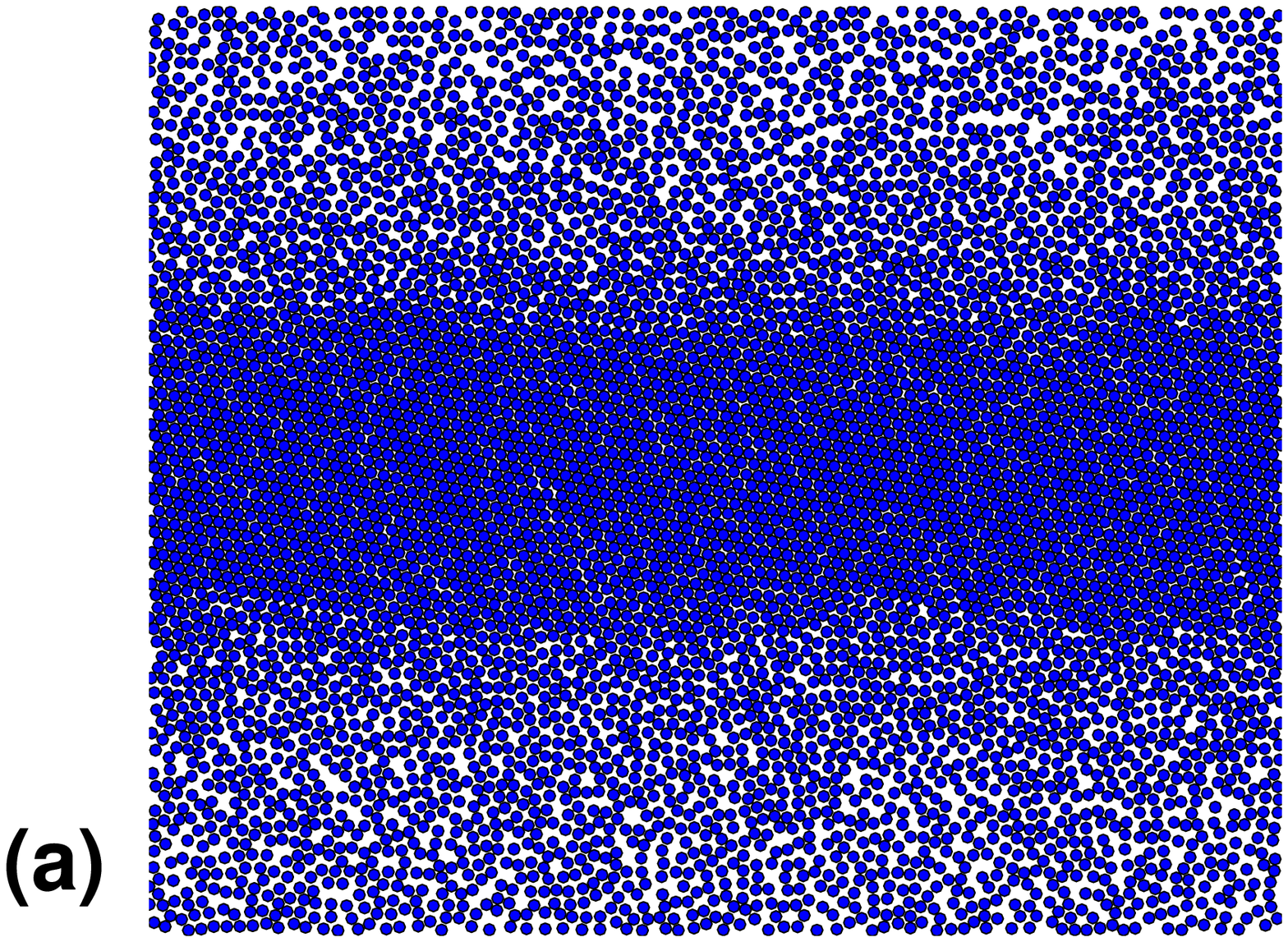}
%\hspace{-0.1cm}
\includegraphics[width=6cm,clip=]{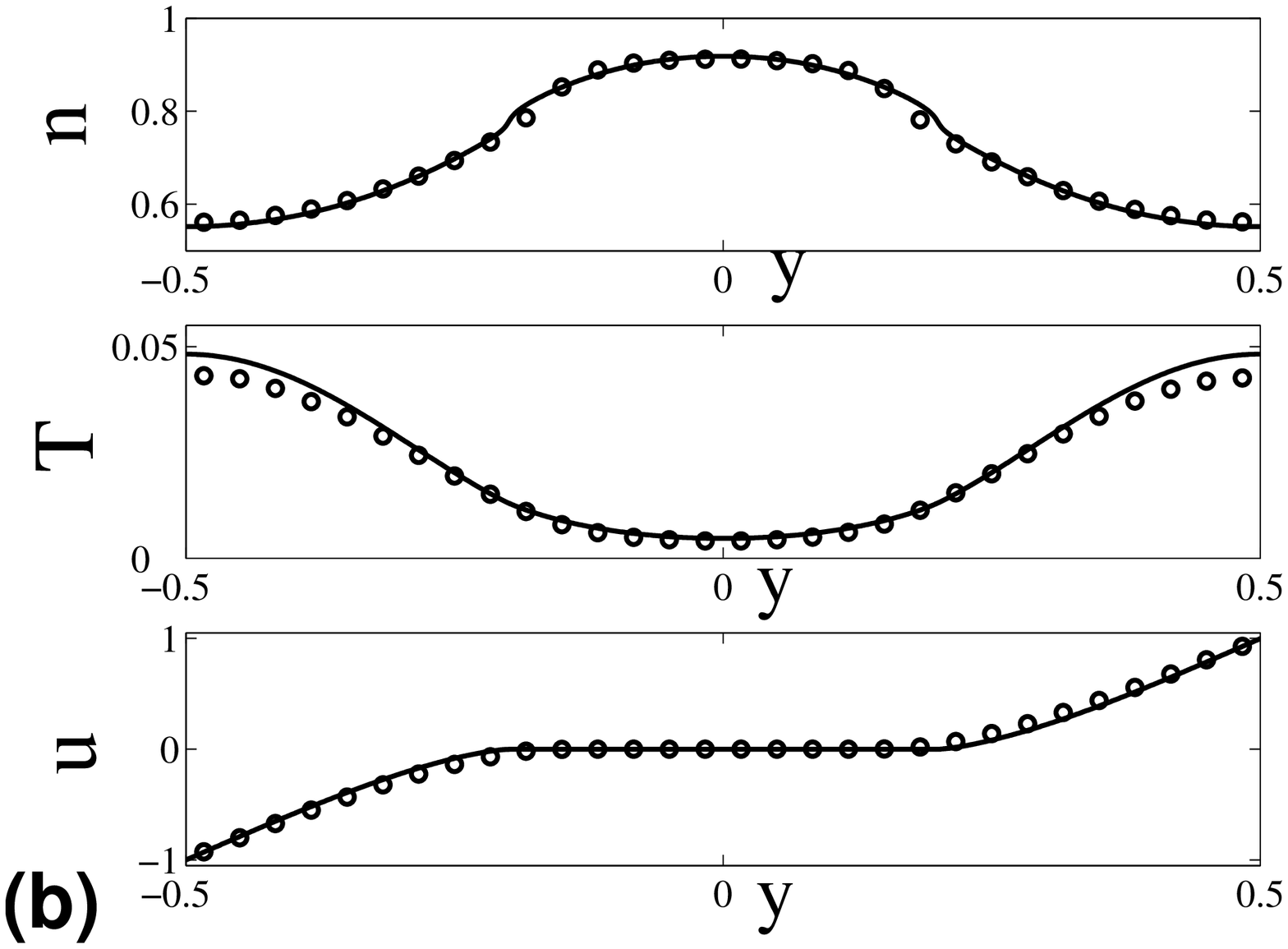}
\caption{A two phase solution. (a): A snapshot of the system from
MD simulations at $t\approx 9\times 10^3$. The solid-like layer in
the middle (which is at rest) is surrounded by two fluid layers.
(b): The corresponding hydrodynamic profiles: theory versus MD
simulations. The density (upper panel), temperature (middle
panel), and velocity (lower panel) profiles in the system, both
from solution of Eqs.~(\ref{steady}) (solid lines) and from MD
simulations (circles). The total number of particles is $N=6480$,
the restitution coefficient is $r=0.99$, the system dimensions are
$L \simeq 97 d$, $H \simeq 80 d$. The corresponding hydrodynamical
parameters are $R=323.86$ and $f = 0.7242$.} \label{snapshot}
\end{figure}

Our MD simulations show that when $R$ is large enough, the system
consists of three layers: an inner solid-like layer and two outer
fluid layers, see example of a snapshot in Fig.~\ref{snapshot}a.
This fluid-solid coexistence has been recently observed in MD
simulations \cite{Luding03,Glasser}. To describe it
hydrodynamically, one needs to solve Eqs.~(\ref{steady}) for each
layer separately, demanding continuity of the density, of the heat
flux, and of the velocity at the interfaces between the layers.
For the solid-like layer, the first equation of
Eqs.~(\ref{steady}) is replaced by $u=const$, whose value should
be found from the overall solution. We find the solution of
Eqs.~(\ref{steady}) using a numerical shooting procedure, similar
to that described in \cite{Khain}. Interestingly, there exists a
family of solutions (missed in \cite{Khain}), which can be
parameterized by the position of density maximum in the system. In
this study we only consider solutions with a symmetric density
profile (however, other solutions were found in MD simulations
\cite{instability} for moderately dense systems). Consider a solid
layer located in the middle of the system, so the density maximum
is at $y=0$ (as in Fig.~\ref{snapshot}a). Figure \ref{snapshot}b
shows the density, temperature, and velocity profiles derived both
from solution of Eqs.~(\ref{steady}) (solid lines) and from MD
simulations (circles). The density (the upper panel) in the
central part of the system (inside the cluster) is larger than the
density of viscosity divergence. This results in $u=0$ plateau in
the middle (the lower panel). An excellent agreement between the
hydrodynamic theory and MD simulations can be seen.

\section{Phase diagram}

\begin{figure}[ht]
%\vspace{0.1cm}
\centerline{\includegraphics[width=6.0cm,clip=]{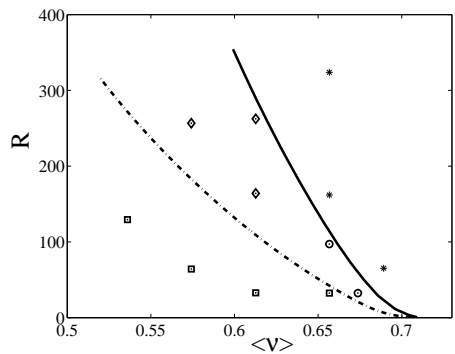}}
\caption{$(R,<\nu>)$ phase diagram. The uniform shear flow is
stable below the dash-dotted line, when $R < R_c(f)$ [see
Eq.~(\ref{lin})]. The two-phase solution is possible above the
solid line, when $R > R_{\ast}(f)$ [where $R_{\ast}(f)$ is
computed from Eqs.~(\ref{steady})]. Symbols denote different
solutions found in MD simulations: uniform shear flow (squares),
two-phase solution (asterisks), nonuniform one-phase solutions
with the density maximum in the middle of the system (rhombuses)
or near the walls (circles). The number of particles in MD
simulations was $N=6450 \pm 30$, and the system height was $H=(79
\pm 1)d$, while the system width $L$ was varied in order to change
the average density in the system.} \label{threshold}
\end{figure}

Now we consider $(R,<\nu>)$ phase diagram, see
Fig.~\ref{threshold}. Let the average density in the system be
smaller than the density of viscosity divergence $n_{\eta}$. If
the hydrodynamic heat loss parameter $R$ exceeds the threshold
value (see the dash-dotted line in Fig.~\ref{threshold})
\begin{equation}
R_c = \pi^2\,f_1 \left[\frac{f_4 (df_3/d\nu)}{(df_4/d\nu)} +
\frac{f_4 f_3 (df_2/d\nu)}{f_2 (df_4/d\nu)} - 2 f_3\right]^{-1}\,,
\label{lin}
\end{equation}
the uniform shear flow becomes unstable (details of linear
stability analysis will be given elsewhere \cite{linear}). Above
this threshold, different solutions with nonuniform density and
temperature profiles are realized. However, the two-phase solution
is possible only when $R$ is larger than some critical value
$R_{\ast}(f)$ (computed from Eqs.~(\ref{steady}), the solid line
in Fig.~\ref{threshold}), so that the density contrast in the
system is sufficiently large, and the maximal density is larger
than $n_{\eta}$. An example of this solution is shown in
Fig.~\ref{snapshot} (which corresponds to the upper asterisk in
Fig.~\ref{threshold}). In order to verify the hydrodynamic
predictions, we performed MD simulations in different regions of
the phase diagram. We found a uniform shear flow solution
(squares) below the dash-dotted line, while above the solid line,
a two-phase solution is realized (asterisks). In the intermediate
region (between the two lines) there are one-phase solutions with
nonuniform density (and temperature) profiles. There are two types
of symmetric (with respect to a density profile) solutions, which
are allowed by Eqs.~(\ref{steady}): the density maximum can be
achieved either in the middle ($y=0$), or near the walls (at
$y=\pm 1/2$). MD simulations show that the system is denser in the
middle when the average area fraction is not large enough
(rhombuses), while for larger area fractions the density is
maximum near the walls (circles).

Interestingly, the solution with solid layer in the middle of the
system seems to be metastable (see also Ref. \cite{Luding03}).
After a sufficiently long time, the cluster starts slowly moving
toward one of the walls until a steady state is reached. We
followed the cluster dynamics by looking on the $y$-component of
center of mass of the system, and on the velocity in the middle
$y=0$. Both quantities reach a plateau at long times, which
corresponds to the position of the cluster near one of the walls.
In this case, the cluster moves as a whole with a velocity which
is slightly lower than the wall velocity.

\section{Summary and discussion}

The dynamics of a dense rapid shear granular flow in two
dimensions is analyzed both theoretically (applying granular
hydrodynamics) and by means of event-driven molecular dynamics
simulations. We presented a general phase diagram and described
different steady-state solutions. The most interesting solution
describes a flow consisting of three layers: an inner solid-like
layer and two outer fluid layers. This solution is possible due to
the fact that viscosity diverges at a lower density than other
constitutive relations \cite{Khain,Luding,BBB}. A possible
direction of future research is increasing the streamwise length
of the system. In this case, the layered structures may give way
to wavy patterns \cite{Glasser}. For moderate densities, these
structures can be explained theoretically \cite{Alam}, while for
higher densities viscosity divergence should be taken into
account.

\begin{acknowledgments}
I am grateful to B. Meerson and I. Aranson for fruitful
discussions. I thank the Michigan Center for Theoretical Physics
for support.
\end{acknowledgments}


\begin{thebibliography}{}
\bibitem{BBB} Note that the assumption that shear viscosity diverges at the
same density as other constitutive relations but more rapidly
\cite{Bocquet} does not agree with MD simulations.
\end{thebibliography}
\end{document}